\documentclass[prl,aps,showpacs,preprintnumbers,twocolumn]{revtex4}
\input{epsf.sty}
\def\beq{\begin{equation}}
\def\eeq{\end{equation}}
\def\eqa{\begin{eqnarray}}
\def\eeqa{\end{eqnarray}}

\newcommand{\roughly}[1]{\mathrel{\raise.3ex\hbox{$#1$\kern-0.85em
\lower1ex\hbox{$\sim$}}}}

\begin{document}
\preprint{DIAS-STP-06-23}
\title{The Quantum Hall Effect in Graphene:\\Emergent Modular Symmetry and the Semi-circle Law}
\author{C.P.~Burgess$^{1,2}$ and Brian P. Dolan$^{3,4}$\\
\vspace{3mm}
{\small $^1$ Department of Physics and Astronomy, McMaster University,\\
\qquad 1280 Main Street West, Hamilton, Ontario, Canada, L8S 4M1.}\\
{\small $^2$ Perimeter Institute, 31
Caroline Street North, Waterloo, Ontario, Canada.}\\
{\small $^3$ Dept. of Mathematical Physics,
National University of Ireland, Maynooth, Ireland.}\\
{\small $^4$ School of Theoretical Physics, Dublin Institute for Advanced Studies,
10 Burlington Rd., Dublin, Ireland.}\\
{\small e-mail: cburgess@perimeterinstitute.ca,
bdolan@thphys.nuim.ie}}

\date{11th December 2006}

\begin{abstract}
Low-energy transport measurements in Quantum Hall systems have
been argued to be governed by emergent modular symmetries whose
predictions are robust against many of the detailed microscopic
dynamics. We propose the recently-observed quantum Hall effect in
graphene as a test of these ideas, and identify to this end a
class of predictions for graphene which would follow from the same
modular arguments. We are led to a suite of predictions for high
mobility samples that differs from those obtained for the
conventional quantum Hall effect in semiconductors, including:
predictions for the locations of the quantum Hall plateaux;
predictions for the positions of critical points on transitions
between plateaux; a selection rule for which plateaux can be
connected by low-temperature transitions; and a semi-circle law
for conductivities traversed during these transitions. Many of
these predictions appear to provide a good description of graphene
measurements performed with intermediate-strength magnetic fields.
\end{abstract}


\maketitle
Recent quantum Hall experiments in single graphene layers
\cite{Novoselov,ZTSK} have produced fascinating results which are
markedly different from the now standard quantum Hall effect in
semiconducting monolayers. In this note we point out that these
experimental results, at least for intermediate values of the
external magnetic field, are compatible with a suggestion, first
made in 1989 \cite{Shapere+Wilczek}, that a particular level 2
sub-group of the modular group might be a symmetry of the
quantum Hall effect. Our main purpose is to identify
the experimental implications of this symmetry, including the
prediction that graphene should be expected to exhibit a
semi-circle law, and to encourage experimentalists to look for these.

The idea that the modular group might be relevant to the quantum
Hall effect in semiconducting heterostructures was explored in
more detail in \cite{Lutken+Ross} where the existence of a group
of symmetries was postulated, acting on the complex conductivity
$\sigma=\sigma_{xy} + i \sigma_{xx}$ according to
\beq
    \sigma\rightarrow \frac{a\sigma + b}{c\sigma
    +d} \,,\label{gammadef}
\eeq
where the integers $a$ through $d$ satisfy the constraint $ad-bc =
1$. This defines the modular group, $\Gamma(1)\approx Sl(2;{\bf
Z})/{\bf Z}_2$. These authors argued that phase structure in the
$\sigma$-plane for high-mobility samples was consistent with the
existence of a symmetry of the above form (at least for samples
with electron spins well-split).  Further progress was made in
\cite{Lutken+RossII,Lutken} where it was noticed that odd
denominator states are singled out provided the integer $c$ is
restricted to be even, defining the level-2 subgroup, $\Gamma_0(2)
\subset \Gamma(1)$. This subgroup differs from the subgroup
$\Gamma_\theta \subset \Gamma(1)$ examined in
ref.~\cite{Shapere+Wilczek}, which can be defined by the condition
that the integers $b$ and $c$ both be even, or both be odd. In
particular $\sigma\rightarrow -{1}/{\sigma}$ is in $\Gamma_\theta$
and this is not a symmetry of the usual quantum Hall effect, there
is no evidence in the experimental data for a fixed point at
$\sigma=i$.

Similar conclusions were reached at much the same time in \cite{KLZ},
where more detailed thinking about the microscopic dynamics led to
the Law of Corresponding States, whose action on filling fractions
implies an action on conductivities which agrees with
$\Gamma_0(2)$ symmetry. Modular transformations, restricted to the
real axis $\sigma_{xx}=0$, generate a $\Gamma_0(2)$ action on
ground state wave-functions, which is also implicit in the work of
Jain and collaborators \cite{Jain}.

A puzzle which these considerations left open was the remarkable
robustness of their experimental implications, which are found to
apply well beyond the expected domain of validity of the
microscopic derivation \cite{NonlinearExp}. This robustness was
clarified by subsequent work \cite{PVD,Nonlinear,Witten}, where it
was argued that symmetries of the form eq.~(\ref{gammadef}) emerge
on very general grounds in the low-energy description of
2-dimensional systems, as a general consequence of
pseudo-particle/vortex duality amongst the charge carriers in the
low-energy effective theory relevant to low-temperature transport
measurements.

In particular, ref.~\cite{PVD} showed the existence of two
equivalence classes of systems, with $\Gamma_0(2)$ emerging as the
symmetry relevant for systems related to fermionic charge carriers 
by a symmetry transformation and
$\Gamma_\theta$ emerging as the symmetry relevant for systems
related to bosonic charge carriers by a symmetry transformation.
Indeed, these two subgroups are related by a conjugation operation
which is equivalent to attaching one unit of statistical gauge
field flux to the relevant pseudo-particles, thereby converting
fermions to bosons and {\it vice versa}.

The power of the modular group was further realised in
\cite{Semicircle} where it was shown that the main experimental
evidence for the Law of Corresponding states could be derived
purely as a consequence of the emergent symmetries. Besides
determining the positions of the possible Hall plateaux as
low-temperature fixed points, the symmetries also govern how
transitions can be made between quantum Hall plateaux as the
magnetic field $B$ or the carrier density $n$ is varied. In
particular, the spectacular semi-circle law for $B$-induced
transitions between quantum Hall plateaux in the low-temperature
limit can be derived assuming only that the flow through the
conductivity plane as temperature is varied should be invariant
under $\Gamma_0(2)$ symmetry (acting as in (\ref{gammadef}))
together with symmetry under particle-hole interchange. The
symmetry also implies that the critical points along these
semicircles occur at universal values of the conductivity
--- for integer transitions at
$\sigma=(k+i)/2$, in units where $e^2/h = 1$, where $k$ is an
integer. Although many such points have been observed in
transitions $\sigma: k\rightarrow k+1$ \cite{ModularSymmetry},
spin effects can modify the precise position of the critical
points \cite{Gamma2,Huang}.

It is the great generality of these arguments which suggests
seeking their test using the new Quantum Hall phenomena recently
found in graphene. Graphene is unusual in that there is an extra
twist in the statistics caused by a Berry phase of $\pi$
\cite{BerryPhase} when particles for each of the two low-energy
bands are circulated around the point of band contact of the other
low-energy band. We can realise this extra phase in the relative
statistics of excitations in the two bands in terms of weakly
interacting pseudo-particles carrying one extra unit of flux of a
statistical gauge field --- this flux would conjugate the relevant
symmetry group from the usual $\Gamma_0(2)$ to $\Gamma_\theta$, as
in \cite{PVD}, leading us to use the latter group to understand
the graphene experiments in \cite{Novoselov,ZTSK}. This suggests
that a hierarchical structure of quantum Hall plateaux, related by
$\Gamma_\theta$, should eventually be found in graphene, given
samples with sufficiently high mobility.

Another complication for graphene at moderate magnetic fields is
the spin degeneracy of the conduction electrons due to the small
Zeeman splitting between electron spin states. We provide details
about how such degeneracies alter the implications of modular
symmetries elsewhere \cite{bilayer}, whose arguments lead us to
consider the modular group $\Gamma_\theta$ acting on $\sigma/2$,
with the factor of $1/2$ arising because of the spin degeneracy.
Note that the symmetry therefore includes the transformation
$\sigma\rightarrow\sigma+4$, which in a fermionic system has the
interpretation of Landau level addition with four-fold degeneracy,
such as would be the usual interpretation for this transformation
in graphene.

The consequences of $\Gamma_\theta$ symmetry for the temperature flow
of the conductivities were given in \cite{PVD}, the resulting temperature flow
diagram is reproduced
here as Figure 1. The predictions for graphene differ from those
given in \cite{PVD}, however, in two ways. First,
because of the spin degeneracy Figure 1 here should be
regarded as plotting the real and
imaginary parts of $\sigma/2$ rather than $\sigma$.

\vtop{ \includegraphics{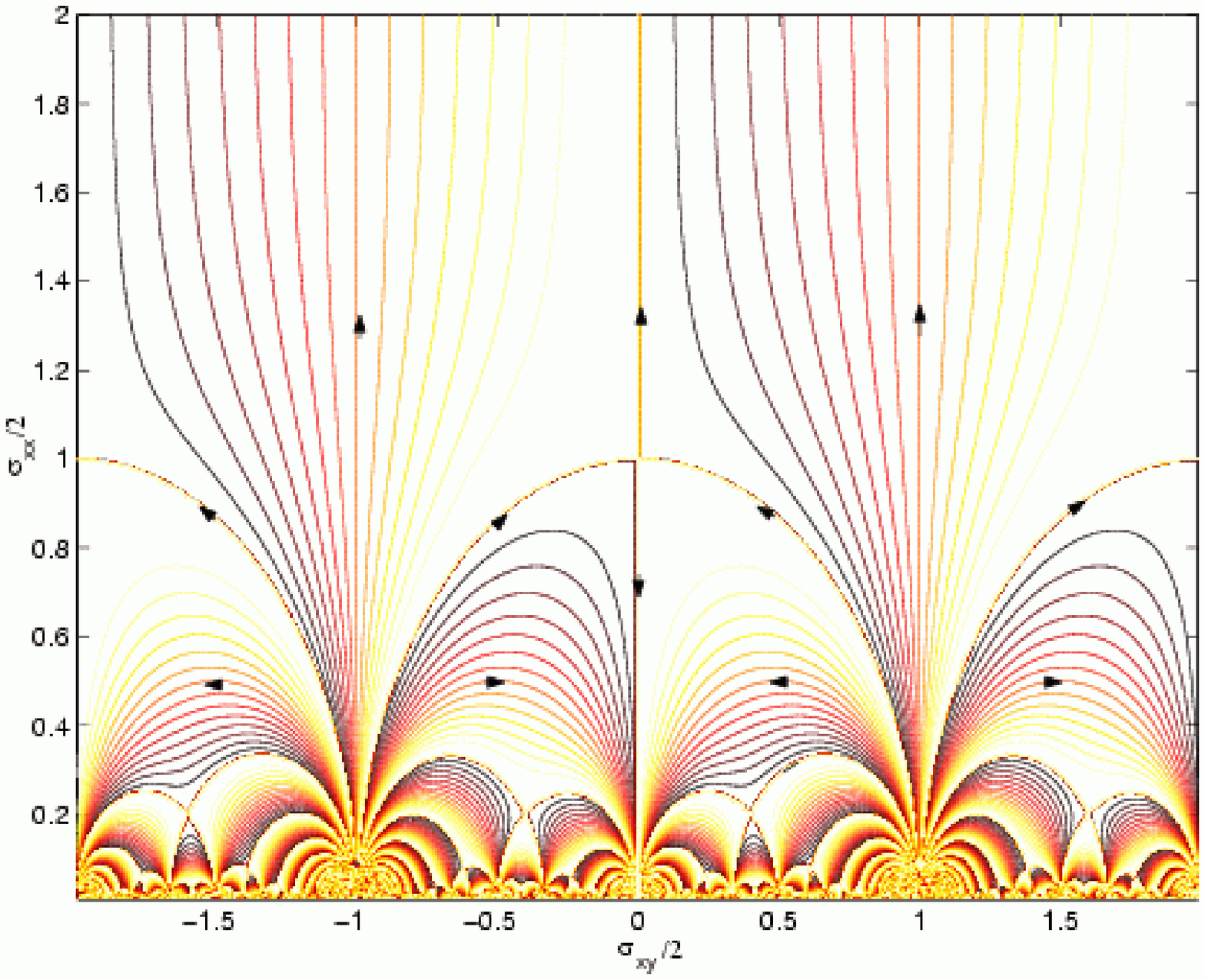} \vskip 5.3cm \centerline{\small
Figure 1: $\Gamma_\theta$ temperature flow (colour online).
} \vskip 0.3cm}

Second, the direction of the temperature flow in the conductivity
plane given in Figure 1 is that given in \cite{PVD}, which was
drawn to represent the direction of decreasing temperature
assuming that the underlying system at zero magnetic field
($\sigma_{xy} = 0$) describes a superconductor/insulator system,
with $\sigma_{xx} \rightarrow i\infty$ or 0 as $T\rightarrow 0$.
(Notice the universal fixed point which is predicted at
$\sigma_{xx} = e^2/h$ describing the superconductor-insulator
transition \cite{Fisher}.) However, graphene is a semi-metal and
so for it one instead expects the conductivity at vanishing
magnetic field to decrease away from $\sigma_{xx} =i\infty$ as $T$ is
lowered. Thus the flow direction shown in Figure 1 should be taken
to represent {\it increasing} temperature for graphene.

We may immediately draw a number of robust consequences concerning
the allowed Hall plateaux for graphene systems, whose validity
relies on the absence of impurities and/or defects which can
destroy the particle-vortex symmetry on which $\Gamma_\theta$ is
based.

\medskip\noindent$\bullet$ {\it Integer Quantum Hall Plateaux:} The
allowed Hall plateaux correspond to the attractive fixed points of
the flow in the limit that $T \to 0$, all of which lie along the
real axis (for which $\sigma_{xx} = 0$. Keeping in mind the
two-fold degeneracy and the flow direction described above,
inspection of the figure (with the arrows reversed) shows that all
such integer fixed points occur at $\sigma_{xy}/2 = 2k+1$, for
integer $k$. This implies integer Quantum Hall plateaux at
$\sigma_{xy} = 4k +2$, as is seen for single-layer graphene
systems \cite{Novoselov,ZTSK} in intermediate magnetic fields.

\medskip\noindent$\bullet$ {\it Fractional Quantum Hall Plateaux:}
The complete set of attractive fixed points as $T \to 0$ occur at
$\sigma_{xy} = 2p/q$, with $p$ and $q$ being relatively prime odd
integers. For higher mobility single-layer graphene samples we
therefore predict the existence of a hierarchy of fractional
Quantum Hall plateaux at precisely these fractional values (and no
others), at least for the same range of magnetic fields for which
the above integer effect is seen.

\medskip

But modular symmetry gives more. Besides identifying the
low-temperature fixed points, taken together with particle-hole
symmetry $\Gamma_\theta$ invariance also dictates the detailed
form of the transitions which can take place between the allowed
plateaux as either magnetic field, $B$, or carrier density, $n$,
are varied. These two symmetries can do so because they suffice to
guarantee \cite{Semicircle} that the {\it caustics} of Figure 1
--- {\it i.e.} the straight lines and semicircles on which those
fixed points lie which are not on the real axis --- are {\it bona
fide} temperature flow lines. (By contrast, the detailed shape of
the remaining flow lines shown in Figure 1 are {\it not} similarly
dictated purely by the symmetries). If one compares a family of
temperature flow curves for different $B$ (or $n$), and connects
the points which correspond to the same value of $T$, then one
finds the resulting curves are inevitably pushed to lie along one
of the caustics in the $\sigma$ plane as $T \to 0$.

This observation leads to the following further predictions:

\medskip\noindent$\bullet$ {\it Selection Rule:} A consequence of
the above arguments is that any low-temperature transition between
a pair of plateaux (with varying $B$ or $n$) must be the image
under $\Gamma_\theta$ of the basic semicircle running from $-1$ to
1 of Figure 1. This implies in particular that transitions are not
possible between arbitrary pairs of Hall plateaux. Instead, the
symmetry allows transitions between two plateaux, $\sigma=2p/q$
and $\sigma=2p'/q'$, only when
\beq
    |pq'-p'q|= 2 \,,\label{SelectionRule}
\eeq
which is compatible with the previous observation that
$p$, $p'$, $q$ and $q'$ are all odd for attractive fixed points. A
selection rule of this type for quantum Hall transitions due to
modular symmetry was first derived in \cite{ModularSymmetry} and
(\ref{SelectionRule}) is the rule for graphene.

\medskip\noindent$\bullet$ {\it Universal Transition Points I:}
Given that the transition between $\sigma = -2$ and $\sigma = 2$
must move along the semicircle of radius 2 centered at the origin,
we have a universal prediction that the value attained by
$\rho_{xx}$ in the absence of Hall conductivity, $\sigma_{xy} =
0$, corresponds to the point $\sigma = 2i$: that is, $\rho_{xx} =
1/2$ (in units where $e^2/h = 1$). Indeed, the experimental data
in \cite{Novoselov} show a peak in $\rho_{xx}$, when
$\sigma_{xy}=0$, very close to $12.9\ k\Omega=h/2e^2$. Modular
symmetry implies that this point should be a fixed point
corresponding to the phase transition which occurs as the system
bifurcates from reaching $\sigma = -2$ to $\sigma = 2$ in the
low-temperature limit. This is the direct analogue of the second
order phase transition at $\sigma={k+i}/2$ in ordinary monolayer
semiconductor samples \cite{Shayegan}. Notice that although these
predictions qualitatively agree with some recent calculations,
which also predict a universal value for $\sigma_{xx}$ when
$\sigma_{xy} = 0$ \cite{GS}, they differ in the universal value
which is predicted.

\medskip\noindent$\bullet$ {\it Universal Transition Points II:}
Since all transition curves are the image of the basic one under
$\Gamma_\theta$, we can predict exactly where the bifurcation
points arise for all other transition curves. Since the group
element required to obtain the transition $\sigma/2: p/q
\rightarrow p'/q'$ from the basic one between $\sigma/2: -1 \to
1$, has $a=(p'-p)/2$, $b=(p'+p)/2$, $c=(q'-q)/2$, and
$d=(q'+q)/2$, it follows that the critical point at $\sigma/2=i$
in the in the $-1$ to $1$ transition is mapped to ${\sigma}/{2} =
({pq + p'q' + 2i})/({q^2+q^{\prime 2}})$ in the $p/q$ to $p'/q'$
transition, and so
\beq
    \sigma=\frac{2(pq+p'q') + 4i}{q^2+q^{\prime 2}} \,.
\eeq
This is therefore the critical bifurcation point for the
transition $\sigma:2p/q\rightarrow 2p'/q'$.

\medskip\noindent$\bullet$ {\it Semi-circle Law:}
Besides predicting the positions of the critical points, as argued
above $\Gamma_\theta$ symmetry determines the entire shape of the
transition curve traced out within the conductivity plane as $B$
or $n$ varies at low temperature, and predicts it to follow a
precisely semi-circular trajectory \cite{Semicircle}. With this
paper we hope to encourage experimentalists to check this
semi-circular prediction.

As a preliminary test we have examined the experimental data in
\cite{Novoselov} (the data in \cite{ZTSK} are not presented in a
way that allowed us to perform the analysis on that experiment), a
figure of which is reproduced here as Figure 2. This figure plots
the Ohmic resistivity, $\rho_{xx}$ and the Hall conductivity,
$\sigma_{xy}$, as reported by \cite{Novoselov} as a function of
carrier density, $n$. Because we were unable to infer directly
from this the trajectory taken in the complex $\sigma$ plane with
sufficient accuracy, we instead chose to directly compare the
curves in this figure with a specific parameterization,
$\sigma(n)$ (related to that in \cite{Holomorphic}, whose detailed form
does not follow purely on symmetry grounds). By assuming
semi-circular trajectories for the transitions $\sigma:
-10\rightarrow -6 \rightarrow -2 \rightarrow 2 \rightarrow 6
\rightarrow 10$ we obtain the blue curves in Figure 2, overlaid on
the experimental curves.
To obtain these curves we use our assumed parameterization to
fit $\rho_{xx}$ for the $\sigma: -2 \to 2$ transition by eye,
constrained by the fact that it should be symmetric about the
vertical axis and that the maximum is not a free parameter but
must lie at $12.9\ k\Omega$. We then derive $\sigma_{xx}$ for this
transition by assuming that it is semi-circle in the
$\sigma$-plane. This semi-circle is then mapped to the other
transitions by using ${\sigma} \to {\sigma} \pm 4$, and
$\rho_{xx}$ is calculated for these transitions. Once $\rho_{xx}$
for the $-2 \rightarrow 2$ transition is determined all of the
other transitions are fit with only one free parameter, namely the
horizontal shift necessary to line up the $\sigma_{xy}$ curve with
the experimental data.
Although our plots use a specific parameterization we believe this
is not likely to be essential for the present purposes --- 
what matters is that the blue curves are semi-circles in the $\sigma$-plane.

\vtop{ \includegraphics{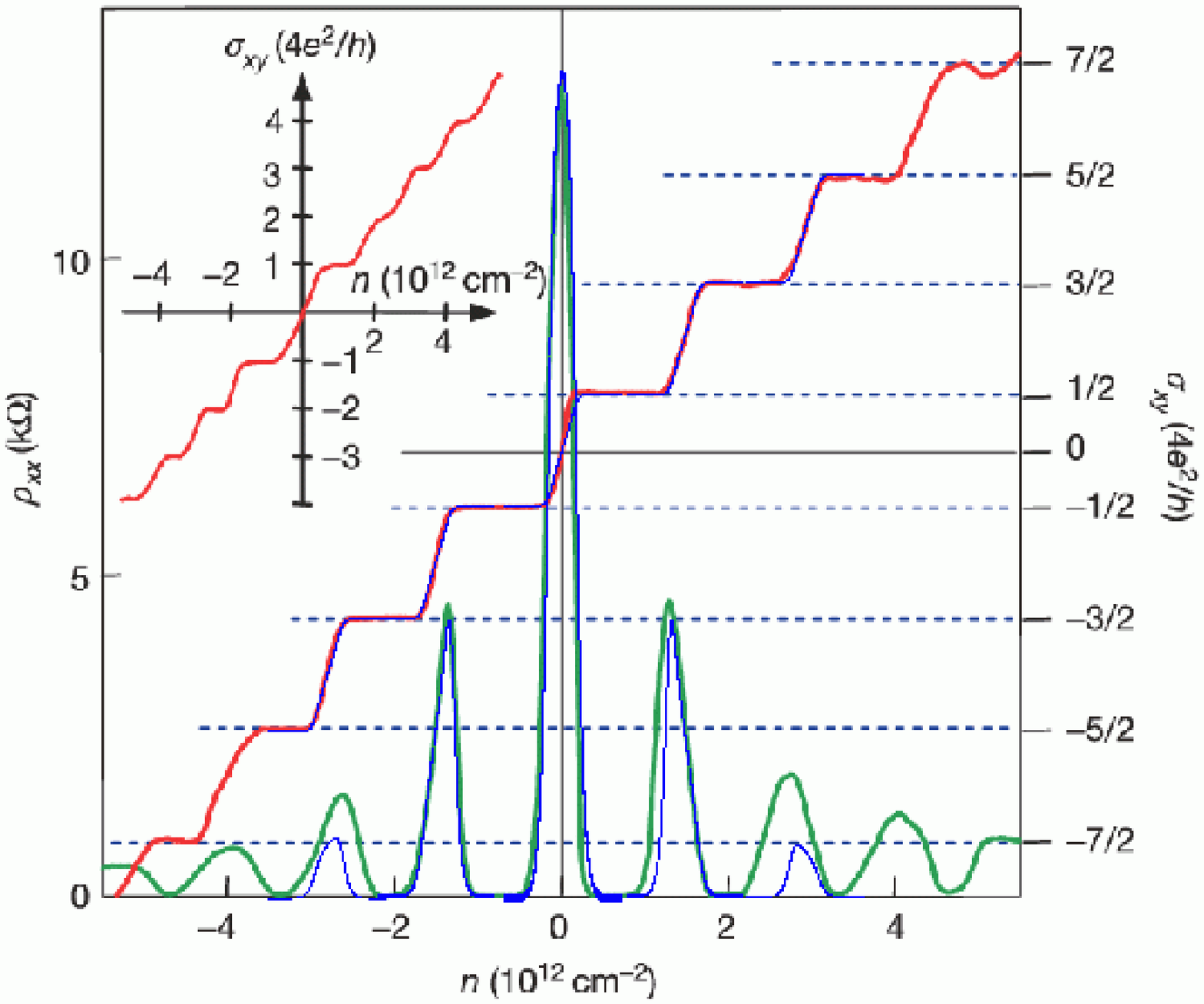} \vskip 7.2cm
\centerline{\small Figure 2: A semicircle (solid blue line) overlaid on
the data}
\centerline{\small of \cite{Novoselov} (colour online). } \vskip
0.2cm }

The visual agreement between the semi-circle curves and the
experimental data seems rather good for $-6\rightarrow -2
\rightarrow 2 \rightarrow 6$, though the fit to $\rho_{xx}$ is
poorer for $2 \rightarrow 6$ than for $-2 \rightarrow -6$, worse
again for the higher plateaux. For the transition $\sigma: 4k - 2
\rightarrow 4k + 2$ a semi-circle implies that the peak in
$\rho_{xx}$ should lie at $\rho_{xx}^{max} = ({2|4k^2-1|})^{-1} \,
k\Omega$, and relative to this the $\pm 10 \rightarrow \pm 6$,
(which a semi-circle predicts should be at $\rho_{xx}=0.85\
k\Omega$) are clearly too high. This indicates a breakdown of
either particle-hole symmetry or the particle-vortex duality
underlying the $\Gamma_\theta$ symmetry.

\medskip

At still higher fields, over $40 T$, the picture changes and more
plateaux are visible, which are interpreted as being due to
lifting the spin degeneracy \cite{HighB}. We do not know how to
interpret these high-field data in terms of modular symmetries, or
whether such an interpretation is possible. These data do
not appear to be compatible with $\Gamma_\theta$ which cannot
exhibit stable plateaux at both even and odd integers (with
pseudo-particles of charge $e$ and $e^2/h=1$) --- perhaps they represent
a transition between two regimes.  It would be very
useful to have data exploring more of the $\sigma$-plane for a
range of values of $B$ and $n$.

The hypothesis that $\Gamma_\theta$ is an underlying symmetry of
the quantum Hall effect in single-layer graphene (for intermediate
magnetic field values) clearly makes many robust predictions.
If borne out by further measurements, the existence of the
emergent $\Gamma_\theta$ symmetry has implications for the
microscopic physics which is responsible. It does so because of
what it says about the properties of the pseudo-particle charge
carriers which appear in the low-energy effective description of
low-temperature transport properties. In particular, the
appearance of the `bosonic' group $\Gamma_\theta$ provides an
additional observational implication of the Berry phase, perhaps
pointing to a composite Fermion picture \cite{CompositeFermions}
for graphene which attaches an odd number of statistical gauge
field flux quanta to each pseudo-particle rather than an even
number.

{\bf Acknowledgments:} BD thanks the Perimeter Institute, for
hospitality during this investigation.

\end{document}